\documentclass[hyper]{JHEP3}
\usepackage{graphicx}
\usepackage{cite}
\usepackage{epsfig,multicol}
\usepackage{epsf}
\usepackage{epstopdf}

\def\q{\end{equation}}
\def\e{\begin{equation}}

\newcommand{\ba}{\begin{eqnarray}}
\newcommand{\ea}{\end{eqnarray}}
\newcommand{\be}{\begin{equation}}
\newcommand{\ee}{\end{equation}}

\title{Resonant Tunneling in Scalar Quantum Field Theory}
\author{S.-H. Henry Tye\footnote{sht5@cornell.edu}~ 
and Daniel Wohns\footnote{dfw9@cornell.edu}
\\ {\em Laboratory for Elementary Particle Physics,
 Cornell University, Ithaca, NY 14853, USA}
 }
 
\date{\today}   

\preprint{}
\abstract{
The resonant tunneling phenomenon is well understood in quantum mechanics. We argue why a similar phenomenon must be present in quantum field theory. We then use the functional
Schr\"odinger method to show how resonant tunneling through multiple barriers takes place in quantum field theory with a single scalar field. We also show how this phenomenon in
scalar quantum field theory can lead to an exponential enhancement of the single-barrier tunneling rate. Our analysis is carried out in the thin-wall approximation. 
}


\begin{document}

\maketitle


\section{Introduction}

In quantum mechanics (QM) the tunneling probability (or the transmission coefficient) of a particle incident on a barrier is typically exponentially suppressed.  
Somewhat surprisingly the addition of a second barrier can increase the tunneling probability for specific values of the particle's energy.  This enhancement in the tunneling
probability, known as resonant tunneling, is due to constructive interference between different quantum paths of the particle through the barriers. Under the right conditions, the tunneling probability can reach unity. This is a very well understood phenomenon in quantum mechanics \cite{Merzbacher}. The first experimental
verification of this phenomenon was the observation of negative differential resistance due to resonant tunneling in semiconductor heterostructures by \cite{CET}. In fact, this
phenomenon has at least one industrial application in the form of resonant tunneling diodes \cite{wiki}.

Tunneling under a single barrier in quantum field theory (QFT) with a single scalar field is well understood, following the work of Coleman and others
\cite{Coleman:1977py,Callan:1977pt}. Despite some arguments given in \cite{Tye:2006tg,Tye:2007ja}, the issue of resonant tunneling in quantum field theory remains open
\cite{Copeland,Saffin:2008vi}. Recently, Sarangi, Shiu and Shlaer suggested that the functional Schr\"odinger method should allow one to study this resonant tunneling phenomenon
\cite{Sarangi:2007jb}. In this paper, we apply this approach to study resonant tunneling in QFT with a single scalar field. We show that resonant tunneling in quantum field theory
does occur and describe its properties. Following Coleman, we shall work in the thin-wall approximation.

\begin{figure}
\begin{center}
\includegraphics[width=0.8\textwidth]{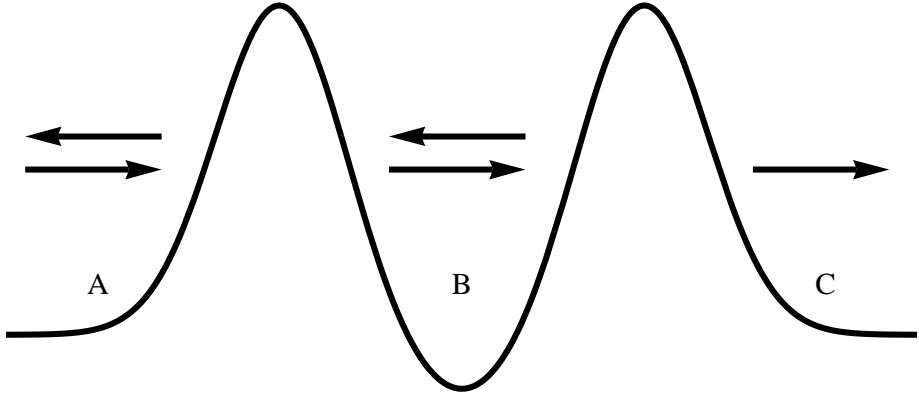}
\caption{A one-dimensional potential in quantum mechanics with three local minima separated by two barriers. Consider an incoming particle from the left. The resonant tunneling effect can take place in the tunneling from $A \to C$ via $B$. With the appropriate energy for the particle, the tunneling probability or transmission coefficient may be as large as unity.}
\label{vacua}
\end{center}
\end{figure}

Before going into any details, it is useful to give an intuitive argument why some effect like resonant tunneling should happen in QFT. Consider the following tunneling process
for a potential shown in Figure \ref{vacua}. Let the tunneling rate from $A \to B$ be $\Gamma_{A \to B} = De^{-S}$, while its tunneling probability $T_{A \to B} =Ke^{-S}$, which
are taken to be exponentially small. The prefactor $D$ or $K$ is of order unity with the proper dimension. Here we shall focus on the exponential factor. Suppose the tunneling
rate from $B \to C$ is given by $\Gamma_{B \to C}$, which is also exponentially suppressed. Both $T_{A \to B}$ and $T_{B \to C}$ are evaluated using standard WKB method. A naive
WKB analysis will suggest that the tunneling from $A \to C$ is doubly exponentially suppressed, i.e., $T_{A \to C} \approx T_{A \to B} T_{B \to C}$. However, this is not correct. Consider the typical time, namely $t_{A \to C}$, it takes to go from $A \to C$. It should be the sum of the time it takes to go from $A \to B$ plus the time it takes to go from $B \to C$. Since the typical time is simply the tunneling (or decay) time, which is the inverse of the rate of tunneling, we have  
\e
\label{tAtoC}
\frac{1}{\Gamma_{A \to C}} = t_{A \to C} = t_{A \to B} + t_{B \to C} = \frac{1}{\Gamma_{A \to B}} + \frac{1}{\Gamma_{B \to C}}  
\q
So it follows that 
\e
\label{gammaAtoC}
T_{A \to C} \approx \frac{T_{A \to B} T_{B \to C}}{T_{A \to B} + T_{B \to C}}
\q
which is clearly not doubly suppressed. For the special case where $T=T_{A \to B} =T_{B \to C}$, we see that $T_{A \to C} \approx T/2$. 

In QM, the exponential enhancement from $T_{A \to C} \approx T^2$ to $T/2$ is due to the resonant tunneling effect. At the resonances,  $T_{A \to C} \approx 1$ while it is $\approx T^2$ off resonances.
For a generic incoming wavefunction, with a spread in energy eigenvalues covering one or more resonances, this resonant effect yields $T_{A \to C} \approx T/2$, so the resonances
typically dominate the tunneling process $A \to C$. More generally, the relations (\ref{tAtoC}) and (\ref{gammaAtoC}) are reproduced \cite{Tye:2006tg,Tye:2007ja}. Since the
argument for (\ref{tAtoC}) is general, it should apply in QFT as well as in QM. This suggests that some phenomenon like resonant tunneling must take place in QFT. The challenge is
to find and understand it.

The functional Schr\"odinger method was developed by Gervais and Sakita \cite{Gervais:1977nv} and Bitar and Chang \cite{Bitar:1978vx}. It starts with the idea that tunneling is
dominated by the most probable escape path (MPEP) developed by Banks, Bender and Wu \cite{Banks:1973ps}. In QFT with a single scalar field $\phi$, this path in the field space is
described by $\phi_0({\bf x}, \lambda)$, where ${\bf x}$ stands for the spatial coordinates and $\lambda$ is a parameter that parametrizes the field configurations in the MPEP. In Coleman's
Euclidean instanton approach, $\lambda$ is chosen to be the Euclidean time $\tau$, and the $O(4)$ symmetry of the instanton simplifies the analysis. On the other hand, the
functional Schr\"odinger method allows one to make a different choice of $\lambda$. In the leading order WKB approximation, a generic choice leads one to a simple
time-independent Schr\"odinger equation. It is not surprising that the resulting WKB formula for a single barrier tunneling process reproduces that of the Euclidean instanton
approach. We shall present a step by step comparison so the equivalence of the two approaches is transparent. It is also obviously clear that the  functional Schr\"odinger method
is cumbersome by comparison. However, this method has the great advantage of being immediately generalizable to the double (actually multiple) barrier case. The underlying reason
is that the same real parameter $\lambda$ parametrizes both the under barrier (Euclidean time $\tau$) and the classically allowed (Minkowski time $t$) regions. 

For tunneling from vacuum $A$ to vacuum $C$ via the intermediate vacuum $B$ in scalar QFT, we consider the simultaneous nucleation of two bubbles, where the outside bubble
separates $A$ from $B$ and the inside one separates $B$ from $C$. The functional Schr\"odinger method reduces this problem, in the leading WKB approximation, to a one-dimensional
time-independent QM problem with $\lambda$ as the coordinate. The resulting double barrier potential in $\lambda$, namely $U(\lambda)$, allows us to borrow the QM analysis to
show the existence of resonant tunneling. In the case when both bubbles grow classically after nucleation, the tunneling process from $A$ to $C$ will be completed. The tunneling
rate is exponentially enhanced compared to the naive case. In the case where the inside bubble classically collapses back after its nucleation (because the bubble is too small for
the difference in the vacuum energies to overcome the surface term due to the domain wall tension), only the tunneling from $A$ to $B$ is completed. Still the tunneling rate from $A$ to $B$ can be
exponentially enhanced. To draw a distinction between these two different phenomena, we call the second process catalyzed tunneling. For catalyzed tunneling, vacuum $C$ plays the role of a catalyst.

The rest of this paper is organized as follows. Resonant tunneling in quantum mechanics is reviewed in Section \ref{QM}. This review follows that in \cite{Merzbacher,Tye:2006tg} . As we shall see, the functional Schr\"odinger method reduces the QFT problem to a QM problem, so the resonant tunneling formalism in QM presented here goes
over directly. In Section \ref{EIM}, we briefly review Coleman's Euclidean action approach, following \cite{Coleman:1977py}. In Section \ref{functional_schroedinger}, we present
the functional Schr\"odinger method.  Here, the discussion follows closely that given by Bitar and Chang \cite{Bitar:1978vx,Chang} and we need only the leading order WKB
approximation. For the single-barrier tunneling process, we see how Coleman's result is reproduced. In Section \ref{RTQFT}, we discuss the double-barrier case. This is the
main section of the paper. Here we find that resonant tunneling can occur in two different ways. It can enhance the tunneling from vacuum $A$ to vacuum $C$ via the intermediate
vacuum $B$, or it can enhance the tunneling from $A$ to $B$ in the presence of $C$.  Section \ref{Conclusion} contains some remarks. The appendix contains a discussion on the particular ansatz for $\phi$ used in the main text.

\section{Review of Resonant Tunneling in Quantum Mechanics} \label{QM}

We first briefly review resonant tunneling in quantum mechanics.  We consider a particle moving under the influence of a one-dimensional potential $V(x)$ with three vacua shown in
Figure \ref{vacua}.  Using the WKB approximation to solve the Schr\"odinger equation $H \psi = E \psi$ for the wavefunction of the particle $\psi(x)$ gives the linearly
independent solutions
\e
\psi_{L,R}(x) \approx \frac{1}{\sqrt{k(x)}} \exp \bigg( \pm i \int dx k(x) \bigg)
\q
in the classically allowed region, where $k(x) = \sqrt{\frac{2m}{\hbar^2}(E-V(x))}$, and
\e
\psi_{\pm}(x) \approx \frac{1}{\sqrt{\kappa(x)}} \exp \bigg( \pm \int dx \kappa(x) \bigg)
\q
in the classically forbidden region, where $\kappa(x) = \sqrt{\frac{2m}{\hbar^2}(V(x)-E)}$.  A complete solution is given by $\psi(x) = \alpha_L \psi_L(x)+\alpha_R \psi_R(x)$ in the classically
allowed region and $\psi(x) = \alpha_+ \psi_+(x)+\alpha_- \psi_-(x)$ in the classically forbidden region.  To find the tunneling probability from A to B we need to determine the
relationship between the coefficients $\alpha_{L,R}$ of the components $\psi_{L,R}$ in vacuum A and the coefficients $\beta_{L,R}$ in vacuum B.  The WKB connection formulae give 
\ba
\label{match}
\left(
\begin{array}{c}
\alpha_R \\
\alpha_L \end{array}
\right) 
=
\frac{1}{2}\left(
\begin{array}{cc}
\Theta + \Theta^{-1}    & i(\Theta - \Theta^{-1}) \\
-i(\Theta - \Theta^{-1})& \Theta + \Theta^{-1} \end{array}
\right)
\left(
\begin{array}{c}
\beta_R \\
\beta_L \end{array}
\right) 
\ea 
where $\Theta$ is given by
\e
\Theta \simeq 2 \exp \left(\frac{1}{\hbar} \int_{x_1}^{x_2} dx \sqrt{2m(V(x) - E)} \right) \,\, ,
\q
and $x_1$ and $x_2$ are the classical turning points.  Setting $\beta_L=0$, the tunneling probablity is given by
\e
\label{T_AtoB}
T_{A \to B} = |\frac{\beta_R}{\alpha_R}|^2 = 4 \left( \Theta + \frac{1}{\Theta} \right)^{-2} \simeq \frac{4}{\Theta ^2} \,\,.
\q
Since $\Theta$ is typically exponentially large, $T_{A \to B}$ is exponentially small.

The same analysis gives the tunneling probability from $A$ to $C$, via $B$, as \cite{Merzbacher,Tye:2006tg},
\e
\label{A_to_C}
T_{A \to C} = 4 \left( \left( \Theta \Phi + \frac{1}{\Theta \Phi}\right)^2 \cos ^2 W + \left( \frac{\Theta}{\Phi} + \frac{\Phi}{\Theta} \right)^2  \sin^2 W \right)^{-1}\,\,,
\q
where
\e
\Phi \simeq 2 \exp \left( \frac{1}{\hbar} \int_ {x_3}^{x_4} dx \sqrt{2m\left(V(x)-E\right)}  \right)   
\q
and
\e
W = \frac{1}{\hbar}\int_{x_2}^{x_3} dx  \sqrt{ 2m(E - V(x))}\,\, , 
\q
with $x_3$ and $x_4$ the turning points on the barrier between B and C.

If $B$ has zero width, $W=0$ so $T_{A \rightarrow C}$ is very small,
\ba
T_{A \rightarrow C} \simeq 4 \Theta^{-2} \Phi ^{-2} =  
T_{A \rightarrow B} T_{B \rightarrow C} /4
\ea
However, if $W$ satisfies the quantization condition for the $n_{B}$th bound states in $B$, 
\ba
W=(n_{B}+1/2) \pi
\label{resonancecond}
\ea
then $\cos W =0$, and the tunneling probability approaches a small but not necessarily 
exponentially small value
\ba
T_{A \rightarrow C}  = \frac{4}{\left(\Theta/\Phi+ \Phi/\Theta \right)^{2}} 
\label{resonancetnn}
\ea
This is the resonance effect. If $T_{A \rightarrow B}$ and $T_{B \rightarrow C}$ are very different, 
we see that $T_{A \rightarrow C}$ is given by the smaller of the ratios between
$T_{A \rightarrow B}$ and $T_{B \rightarrow C}$. 
Suppose $T_{A \rightarrow B} \rightarrow T_{B \rightarrow C}$.
Following (\ref{resonancetnn}), we see that $T_{A \rightarrow C}   \rightarrow  1$ that is, the tunneling probability approaches unity.  Notice that the existence of resonant tunneling effect
here is independent of the detailed values of $\Theta$, $\Phi$, and $W$. 

The above phenomenon is easy to understand in the Feynman path integral formalism. A typical tunneling path starts at $A$ and tunnels to $B$. It bounces back and forth $k$ times,
where $k=0,1,2,... \infty$, before tunneling to $C$. When the Bohr-Sommerfeld quantization condition (\ref{resonancecond}) is satisfied, all these paths interfere coherently,
leading to the resulting resonant tunneling.   On the other hand, if we raise the energy of the local minimum at B above the incoming energy $E$, then $W=0$ and
$T_{A \rightarrow C} \sim T_{A \rightarrow B} T_{B \rightarrow C}$ which is typically doubly exponentially suppressed.

\section{Coleman's Euclidean Instanton Method} \label{EIM}

Single-barrier tunneling in quantum field theory was studied by \cite{Coleman:1977py} using the Euclidean instanton method.
For concreteness, we will focus on the $(3+1)$-dimensional scalar field theory in Minkowski space described by the Lagrangian
\e
\mathcal{L}= \frac{1}{2} \dot{\phi}^2 - \frac{1}{2} (\nabla \phi)^2 - V(\phi)
\q
with an asymmetric double well potential
\e
\label{potential}
V(\phi) = \frac{1}{4} g (\phi ^ 2 - c ^ 2) ^ 2 - B (\phi + c)
\q
with $B$ a small symmetry-breaking parameter.  The potential at the false vacuum at $\phi=-c$ is zero, while the potential at the true vacuum at  $\phi=+c$ is $-\epsilon \approx
-2 B c$. The energy density difference between the two minima, namely $\epsilon$, is assumed to be small so that we may confine our analysis of this potential to the thin-wall
regime. In the under-barrier (i.e., classically forbidden) region, one starts with the Euclidean action $S_E(\phi (\tau,{\bf x}))$ where $\tau$ is the Euclidean time. Solving the
resulting Euclidean equation of motion for $\phi$ (with appropriate boundary conditions) and substituting it back into  $S_E(\phi (\tau,{\bf x}))$ yields $S_E$. In the semi-classical
limit \cite{Coleman:1977py} showed that the tunneling rate per unit volume is 
\e
\Gamma / V = A \exp(-S_E / \hbar) \,\,,
\q 
where the subexponential prefactor $A$ studied in \cite{Callan:1977pt} will be unimportant for our purposes.
The solution $\phi (\tau, {\bf x}, R)$ to the Euclidean equation of motion is the familiar $O(4)$-symmetric domain-wall solution
\e
\label{phidw}
\phi_{DW} (\tau, {\bf x}, R) = - c \tanh \bigg(\frac{\mu}{2} (r - R)\bigg) \,\,,
\q
where $\mu$ measures the (inverse) thickness of the domain wall,
\e
\mu = \sqrt{2 g c ^2 }
\q
so a relatively large $\mu$ (i.e., thin wall) is assumed. Here
\e
r^2 = {|{\bf x}|}^2 +\tau^2,
\q
and $R$ is the radius of the bubble. The bubble wall sits at $r=R$, so $\phi=+c$ for $r \ll R$ and $\phi=-c$ for $r \gg R$. That is, the bubble is surrounded by the false vacuum.
It is useful to introduce the tension $S_1$ of the domain wall,
\e
 S_1 =  \int^c_{-c} d\phi \sqrt{2 V(\phi)} 
    \approx \int^c_{-c} d\phi \sqrt{\frac{g}{2}(\phi^2 - c^2)^2}
    = \frac{2}{3}\mu c
\q
so the Euclidean action of this solution is now given by
\e
S_E = - \frac{1}{2} \pi^2 R^4 \epsilon + 2\pi ^2 R^3 S_1 \,\,.
\q
where the first term is the four-volume times the energy density difference $\epsilon$ while the second term is the contribution of the domain wall.
Setting the variation of this $S_E$ to zero yields
\e
\label{zeroE}
\mathcal{E} =  - \frac{4}{3} \pi R^3 \epsilon + 4\pi R^2 S_1 =0 \,\,.
\q
So the action is stationary for 
\e\label{Rlambdac}
R = \lambda_c = 3 S_1/\epsilon,
\q
which gives us
\e
\label{S_E0}
S_E =  \frac{\pi^2}{2} S_1 \lambda_c^3 = \frac{27 \pi^2}{2} \frac{S_1^4}{\epsilon^{3}}
\q
What happens to the bubble after nucleation? The bubble will behave in a way to decrease the energy $\mathcal{E}$, i.e., $d\mathcal{E}/dR<0$. It is easy to see that the bubble prefers to grow (classically) as long as
\e\label{Rgrow}
R > 2\lambda_c/3
\q
which is the case here. So,
once the bubble is created with radius $\lambda_c$, the domain wall starts at rest and moves (classically) outwards, eventually attaining relativistic speed.
Notice that the condition (\ref{zeroE}) is simply the classical energy conservation equation: at the moment right after bubble nucleation, the total energy $\mathcal{E}$ of the bubble and its interior equals to that of the original false vacuum in the region, which is zero.

\section{Functional Schr\"odinger Method} 
\label{functional_schroedinger}

Now let us introduce the functional Schr\"odinger method and apply it to the single barrier tunneling process discussed above in the same scalar QFT. 
In the semi-classical regime, a discrete set of classical paths, namely the most probable escape paths (MPEP) in configuration space give the dominant contributions to the vacuum tunneling rate \cite{Banks:1973ps,Gervais:1977nv,Bitar:1978vx}.  Essentially this approximation allows us to reduce an infinite-dimensional quantum field theory calculation to a one-dimensional
quantum mechanical computation.  The effects of nearby paths can be included systematically in an $\hbar$ expansion, and were calculated to $O(\hbar^2)$ in \cite{Gervais:1977nv,Bitar:1978vx},
but will not be relevant for the rest of our analysis.  

The Hamiltonian for a scalar field $\phi(t,{\bf x})$ where ${\bf x}$ denotes the three spatial directions is
\e
H = \int d^3{\bf x} \bigg(\frac{\dot{\phi}^2}{2}+ \frac{1}{2} (\nabla \phi)^2 +V(\phi)\bigg) \,\,.
\q
where $V(\phi)$ is that given in (\ref{potential}).
To quantize the field theory we use $[\dot{\phi}({\bf x}),\phi({\bf x}')]=i \hbar \delta^3({\bf x}-{\bf x}')$ to replace $\dot{\phi}$ with $-i\hbar \delta / \delta \phi$.  This
replacement allows us to write the time-independent functional Schr\"odinger equation as
\e
\label{func_SE}
H \Psi(\phi({\bf x})) = E \Psi(\phi({\bf x}))
\q
where
\e
H = \int d^3{\bf x} \bigg(- \frac{{\hbar}^2}{2}\bigg( \frac{\delta}{\delta \phi ({\bf x})}\bigg)^2+ \frac{1}{2} (\nabla \phi)^2 +V(\phi)\bigg) \,\,,
\q
and the eigenvalue $E$ is the energy of the system.
As usual $\Psi(\phi({\bf x}))$ is the amplitude that gives a measure of the likelihood of the occurance of the field configuration $\phi({\bf x})$. 

With the ansatz $\Psi(\phi) = A \exp (-\frac{i}{\hbar} S(\phi))$ where $A$ is constant, the functional Schr\"odinger equation (\ref{func_SE}) becomes
\e
\label{func_SE_S}
\int d^3{\bf x} \left(  -\frac{\hbar^2}{2} \left[ \frac{i}{\hbar}\frac{\delta^2 S(\phi)}{\delta \phi^2} - \frac{1}{\hbar^2}\left( \frac{\delta S(\phi)}{\delta \phi}\right)^2 \right]+
\frac{1}{2}(\nabla \phi)^2 + V(\phi) \right) e^{\frac{i}{\hbar}S(\phi)} = E e^{\frac{i}{\hbar}S(\phi)}\,\,.
\q
Expanding $S(\phi)$ in powers of $\hbar$, $S(\phi) = S_{(0)}(\phi) + \hbar S_{(1)}(\phi) + ...$, and comparing terms with equal powers of $\hbar$ the functional Schr\"odinger
equation
(\ref{func_SE_S}) yields
\ba
\label{semiclassical_expansion}
\int d^3{\bf x} \left[ \frac{1}{2}\left(\frac{\delta S_{(0)}(\phi)}{\delta \phi}\right)^2 + \frac{1}{2}(\nabla \phi)^2 + V(\phi) \right] = E, \\ \nonumber
\int d^3{\bf x} \left[ -i\frac{\delta^2 S_{(0)}(\phi)}{\delta \phi^2} + 2 \frac{\delta S_{(0)}(\phi)}{\delta \phi}\frac{\delta S_{(1)}(\phi)}{\delta \phi} \right] = 0,\\ \nonumber
etc.
\ea
The infinite set of nonlinear equations (\ref{semiclassical_expansion}) on an infinite-dimensional configuration space can be reduced to a one-dimensional equation in the leading
approximation.  For our purpose here, we shall focus on $S_{(0)}$ and ignore the higher-order corrections $S_{(1)}$, $S_{(2)}$, etc.  The essential idea 
is that there is a trajectory in the configuration space of $\phi({\bf x})$, known as the most probable escape path (MPEP),
perpendicular to which the variation of $S_{(0)}$ vanishes, and along which the variation of $S_{(0)}$ is nonvanishing.  We use $\lambda$ to parametrize this path in the configuration space of $\phi(x)$, so the MPEP is $\phi ({\bf x}, \lambda)$.
This MPEP satisfies
\ba
\label{MPEP1}
\frac{\delta S_{(0)}}{\delta \phi_{||}} \mid_{\phi_0 ({\bf x},\lambda)} &=& 
C(\lambda) \frac{\partial \phi_0}{\partial \lambda}, \nonumber \\
\frac{\delta S_{(0)}}{\delta \phi_{\perp}} \mid_{\phi_0 ({\bf x},\lambda)} &=& 0\,\,.
\ea
Along the MPEP, we have 
\e
\label{MPEP2}
 \frac{\partial S_{(0)}}{\partial \lambda} = \int d^3{\bf x} \frac{\partial \phi_0 ({\bf x},\lambda)}{\partial \lambda} \frac{\delta S_{(0)}}{\delta \phi_{||}}
 \mid_{\phi_0 ({\bf x},\lambda)}
 \q
 so $C(\lambda)$ is determined and we have
\e
\label{MPEP}
\frac{\delta S_{(0)}}{\delta \phi_{||}} \mid_{\phi_0({\bf x},\lambda)} = \frac{\partial S_{(0)}}{\partial \lambda} \left( \int d^3{\bf x} \left[\frac{\partial \phi_0({\bf x},\lambda)}{\partial \lambda}\right]^2\right)^{-1} \frac{\partial \phi_0 ({\bf x},\lambda)}{\partial \lambda}, \
\q
We now define the effective potential $U(\lambda)=U(\phi({\bf x},\lambda))$, so
\e
\label{Upot}
U(\phi({\bf x},\lambda)) = \int d^3{\bf x} \left( \frac{1}{2}(\nabla \phi({\bf x},\lambda))^2 + V(\phi({\bf x},\lambda)) \right) \,\,,
\q
then the classically allowed regions have $U(\phi_0({\bf x}, \lambda)) < E$ and the classically forbidden regions have $U(\phi_0({\bf x}, \lambda)) > E$.  

Using the zeroth-order equation in (\ref{semiclassical_expansion}) with (\ref{MPEP}) we find the WKB equation
\e
\label{WKB}
-\frac{1}{2}\left( \int d^3{\bf x} \left[\frac{\partial \phi_0}{\partial \lambda}\right]^2\right)^{-1} \left( \frac{\partial S_{(0)}}{\partial \lambda} \right)^2
= U(\phi({\bf x},\lambda)) - E\,\,.
\q
To find the WKB wavefunctional it is sometimes useful to rewrite (\ref{WKB}) in terms of the path length $ds$ in the configuration space.  This path length is defined by
\e
\label{ds}
(ds)^2 = \int d^3{\bf x} (d \phi({\bf x}))^2 = (d\lambda)^2 \int d^3{\bf x} \left( \frac{\partial \phi({\bf x},\lambda)}{\partial\lambda} \right)^2 =(d\lambda)^2 m(\phi({\bf x},\lambda))\,\,.
\q
That is, choosing $s$ to parametrize the MPEP, 
(\ref{WKB}) simplifies to 
\e
\label{WKB_path0}
-\frac{1}{2}\left( \frac{\partial S_{(0)}}{\partial s} \right)^2 = U(\phi({\bf x},s)) - E \,\,.
\q
Now let us consider first the classically forbidden region, $U(\phi({\bf x},s)) >  E$.
The solution to (\ref{WKB_path0}) is
\e
\label{WKB_solution}
S_{(0)} = i \int_0^s ds' \sqrt{2[U(\phi({\bf x},s'))-E]}= i \int_{\lambda_{t1}}^{\lambda_{t2}} d\lambda \bigg(\frac{ds}{d\lambda}\bigg) \sqrt{2[U(\phi({\bf x},\lambda))-E]}\,\,. 
\q
where $\lambda_{t1}$ and $\lambda_{t2}$ are the turning points.
Treating both $\frac{ds}{d\lambda} (\phi ({\bf x},\lambda))$ (\ref{ds}) and $U(\phi({\bf x},\lambda))$ (\ref{Upot}) as functionals of $\phi$, the Euler-Lagrange equation for $\phi({\bf x},
\lambda)$ follows from setting the variation of $S_{(0)}$ to zero.
It turns out that the resulting equation of motion derived from (\ref{WKB_solution}) simplifies considerably if we choose $\tau$ as the parameter where 
\e
\frac{ds}{d\tau} = \sqrt{2[U(\phi({\bf x},\tau))-E]}\,\,.
\q
With $\tau$ as the parameter, setting the variation of (\ref{WKB_solution}) equal to zero now yields
\e
\label{EL}
\frac{\partial^2 \phi({\bf x},\tau)}{\partial \tau^2} + \nabla^2 \phi({\bf x},\tau) - \frac{\partial V(\phi({\bf x},\tau))}{\partial \phi} = 0
\q
Here $\tau$ simply plays the role of Euclidean time, and (\ref{EL}) is simply the Euclidean equation of motion for $\phi({\bf x}, \tau)$. Once we obtain the solution for $\phi({\bf x}, \tau)$, we insert this solution into (\ref{WKB_solution}) to obtain the value $S_0$ for $S_{(0)}$.
So we see that the functional Schr\"odinger method leads both to a determination of $S_0$ along the MPEP and an equation that determines MPEP, namely, $\phi_0({\bf x}, \tau)$ itself.
The subscript ``0" indicates that it is the MPEP in the leading WKB approximation. 
The equation (\ref{EL}) has $O(4)$ symmetry, so it reproduces the familiar $O(4)$ symmetric domain-wall solution (\ref{phidw}) in the thin-wall approximation ($r$ is the
four-dimensional radial coordinate, $r^2 = \tau^2 + |{\bf x}|^2$),
\e
\label{phixtau}
\phi_0 ({\bf x}, \tau) = - c \tanh \bigg(\frac{\mu}{2} (r - \lambda_c)\bigg)
\q
after imposing the boundary conditions, $\phi \to -c$ as $r \to \infty$ and $\phi \to +c$ as $r \to 0$ and $\frac{d\phi(0)}{dr}=0$. Recall that $\lambda_c$ is the critical radius of the bubble given by (\ref{Rlambdac}).

\begin{figure}
\begin{center}
\includegraphics[width=0.8\textwidth]{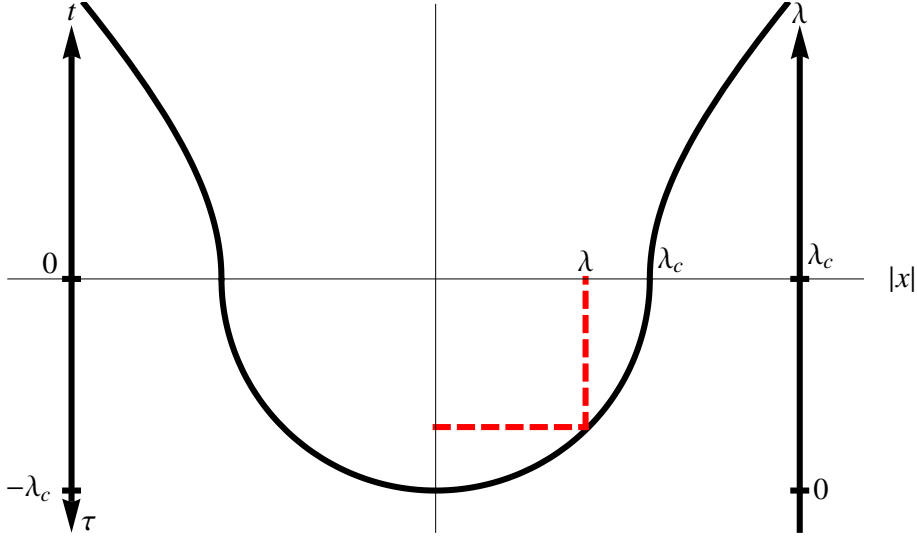}
\caption{Spatial slices of the familiar domain-wall solution (\protect\ref{phi_approx}) can be parameterized by the spatial radius of the bubble of true vacuum, $\lambda$. Here
$\lambda$ is the length of the horizontal dashed line. The radius $\lambda_c$ of the (bottom half) bubble is related to the Euclidean time $\tau$ via $\lambda_c^2= \tau^2 +
\lambda^2$. The choice of the $\lambda$ parameter allows us to go smoothly from the classically forbidden region (below the x-axis) to the classically allowed region (above
the x-axis). }
\label{lambda_plot}
\end{center}
\end{figure}

\begin{figure}
\begin{center}
\includegraphics[width=0.8\textwidth]{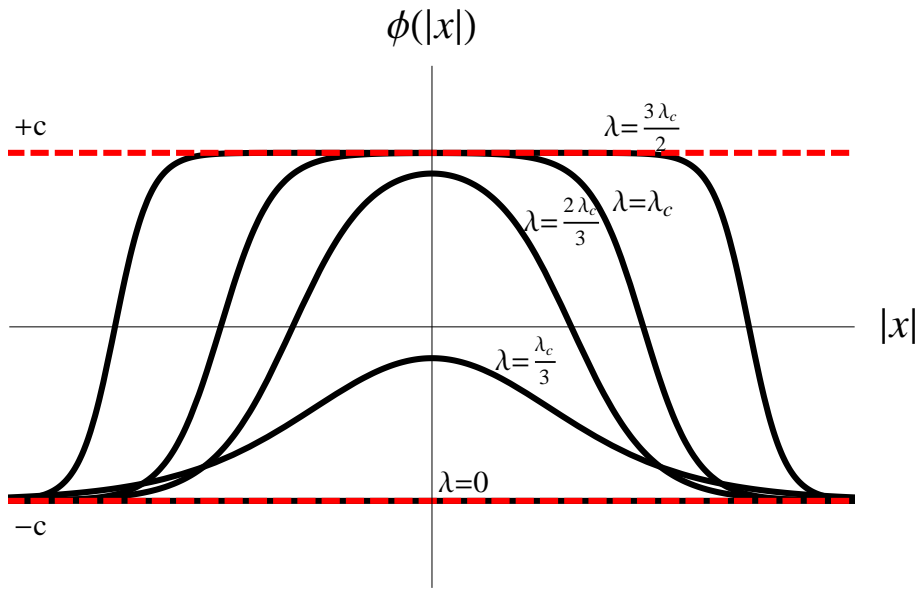}
\caption{The MPEP is a path through field configuration space parameterized by $\lambda$. Here, $x$ is the spatial radius, $\phi= - c$ is the false vacuum and $\phi=+c$ is the true vacuum. For the MPEP given by (\protect\ref{phi_approx}): 
$\lambda \le 0$ corresponds to the false vacuum, $\lambda_c \ge \lambda >0$ corresponds to the formation of the nucleation bubble during the tunneling process,
$\lambda=\lambda_c$ corresponds to the completion of the nucleation of the bubble of true vacuum, and $\lambda \ge \lambda_c$ corresponds to the classical growth of the bubble.}
\label{MPEP_plot}
\end{center}
\end{figure}

For our purpose, it is now convenient to introduce the parameter $\lambda$ 
\e
\lambda = \sqrt{\lambda_c^2 - \tau^2}
\q
which is the spatial radius of the bubble as shown in Figure \ref{lambda_plot}. Here $\phi_0({\bf x}, \lambda) =0$ for $\lambda<0$.
For $\lambda_c > \lambda > 0$ and near the domain wall at $r= \sqrt{{|{\bf x}|}^2+\tau^2}$, we have $r- \lambda_c \approx \pm ({|{\bf x}|}^2 - \lambda^2) / (2 \lambda_c) \approx \pm ({|{\bf x}|} \pm \lambda) \lambda / \lambda_c$.  The
corrections introduced by this approximation are exponentially suppressed far from the domain wall, so we can express the $O(4)$ symmetric solution (\ref{phixtau}) as,
\e
\label{phi_approx}
\phi_0({\bf x}, \lambda) 
 \approx - c \tanh \bigg( \frac{\mu}{2} ({|{\bf x}|} - \lambda) \frac{\lambda}{\lambda_c} \bigg)  \,\,,
\q
This MPEP solution is plotted in Figure \ref{MPEP_plot}. (Slight care is needed for $\lambda \approx 0$, in which case, we simply go back to (\ref{phixtau}).)  For $\lambda<0$ the
solution is the false vacuum $\phi_0 ({\bf x}, \lambda)=-c$ . As $\lambda>0$ increases, quantum fluctuation tends to fluctuate towards the true vacuum. At $\lambda=\lambda_c$, the bubble is created, which then evolves classically for $\lambda > \lambda_c$. 

\begin{figure}
\begin{center}
\includegraphics[width=0.8\textwidth]{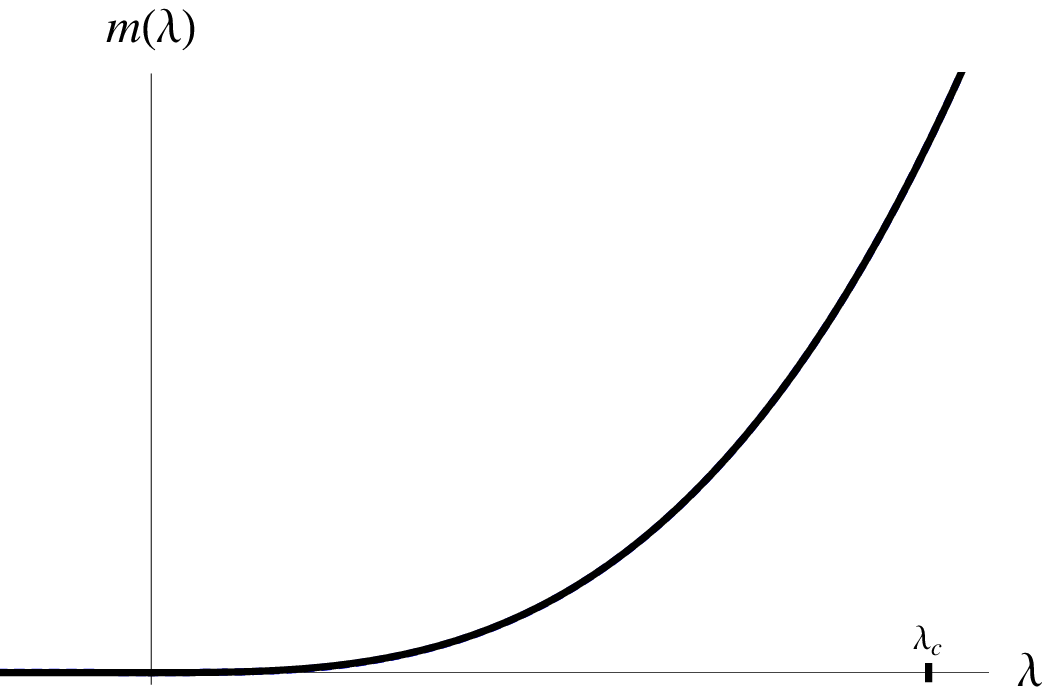}
\caption{The position-dependent mass $m(\lambda)$ for $g = 1$, $c = 1.5$, and $B = 0.1$.  The exact curve is indistiguishable from the approximate curve (\protect\ref{m}) in this plot.}
\label{m_plot}
\end{center}
\end{figure}

\begin{figure}
\begin{center}
\includegraphics[width=0.8\textwidth]{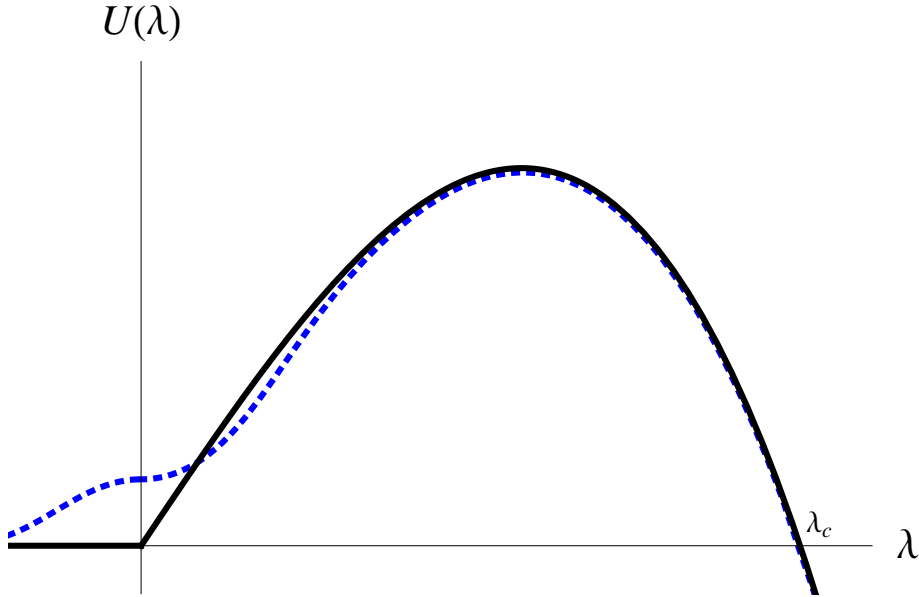}
\caption{The effective tunneling potential $U(\lambda)$ for $g = 1$, $c = 1.5$, and $B = 0.1$.  The black solid curve shows the approximate form of $U(\lambda)$ given in (\protect\ref{U_single}). The blue dotted curve shows the numerical result. The difference is not important for our analysis.}
\label{single_barrier}
\end{center}
\end{figure}

We have effectively reduced the WKB wavefunctional in the classically forbidden region to a WKB wavefunction which can be written using (\ref{WKB_solution}) as
\e
\label{S_0s}
\Psi(\phi({\bf x},\lambda)) = A e^{iS_0/\hbar} =A \exp \left( -\frac{1}{\hbar}\left[\int_{0}^{\lambda_c} d\lambda \sqrt{2m(\lambda)[U(\lambda)-E]}\right] \right) \,\,,
\q
where $m(\lambda)$ is obtained by substituting the MPEP $\phi_0({\bf x}, \lambda)$ (\ref{phi_approx}) into $m(\phi({\bf x}, \lambda))$,
\e
\label{m}
m(\lambda) \equiv \int d^3x \bigg( \frac{\partial \phi_0 ({\bf x}, \lambda)}{\partial \lambda}\bigg) ^2 
     \approx 4 \pi S_1 \frac{\lambda^3}{\lambda_c}  
\q
Here $m(\lambda)$ is the effective mass, which is manifestly positive, and the second equality is obtained in the thin-wall approximation.
It is also straightforward to evaluate $U(\lambda)$ by substituting MPEP $\phi_0({\bf x}, \lambda)$ (\ref{phi_approx}) into $U(\phi({\bf x},\lambda))$ (\ref{Upot}),
\e
\label{U_single}
U(\lambda) \approx \frac{2 \pi S_1}{\lambda_c} \lambda ( \lambda_c^2 - \lambda^2 )	  
\q
This approximation for $U(\lambda)$ is plotted in Figure \ref{single_barrier}. The classically forbidden region for zero-energy tunneling is $0 < \lambda < \lambda_c$. As
mentioned above, $\phi_0({\bf x}, \lambda)$ (\ref{phi_approx}) needs correction for $\lambda \sim 0$. The more accurate $U(\lambda)$ (the blue dotted curve) is also shown in Figure
\ref{single_barrier}. However, this difference is not important for our analysis here.

Given $m(\lambda)$ (\ref{m}) and $U(\lambda)$ (\ref{U_single}), the tunneling problem in QFT has been reduced to a one-dimensional time-independent QM problem.  We can now perform
the integral in the exponent in (\ref{S_0s}) and obtain, for $E=0$,
\e
-iS_0= \frac{27 \pi^2}{4} \frac{S_1^4}{\epsilon^{3}}=S_E/2
\q
which reproduces (\ref{S_E0}). (The factor of two difference is because, here, we are evaluating the exponent of the tunneling amplitude instead of the tunneling rate.)
This result is completely expected since the expression (\ref{S_E0}) is obtained by integrating the Langrangian density with the $O(4)$ symmetric solution (\ref{phixtau}) or (\ref{phidw})
in polar coordinates while here we are performing the same integral, first in the spatial coordinates $x$ and then in the $\lambda$ (or equivalently in $\tau$) coordinate. For an
$O(4)$ symmetric solution, the latter approach is unnecessarily cumbersome.
  
So, before moving on, let us give a preview of the advantage of the functional Schr\"odinger method. 
This will also shed light on the underlying physics. 
In the classically allowed regions, similar arguments lead to the following $S_{(0)}$,
\e
S_{(0)}(\phi({\bf x},\lambda)) =  \int  d\lambda \sqrt{2m(\phi({\bf x},\lambda))[E -U(\phi({\bf x},\lambda))]}
\q
Similar to the previous case, the Euler-Lagrange equation of motion for $\phi$ simplifies if we choose the parameter $t$ such that
\e
\frac{ds}{dt} = \sqrt{2[E-U(\phi({\bf x},t))]}\,\,.
\q
so setting the variation of $S_{(0)}$ with respect to $\phi({\bf x},t)$ to zero leads to 
\e
\label{LEOM}
\frac{\partial^2 \phi({\bf x},t)}{\partial t^2} - \nabla^2 \phi({\bf x},t) + \frac{\partial V(\phi({\bf x},t))}{\partial \phi} = 0\,\,,
\q
where $t$ is simply the normal time, and this is simply the equation of motion for $\phi ({\bf x},t)$ in Minkowski space, $t \ge 0$. Now, instead of $t$, let us choose $\lambda$ as the parameter, where
\e
\lambda = \sqrt{\lambda_c^2 + t^2}
\q
which gives us ($\dot\lambda =d\lambda/dt$),
\e
\frac{\lambda}{\lambda_c} = \frac{1}{\sqrt{1- {\dot\lambda}^2}}
\q
which is simply the Lorentz factor.
Substituting this into the path (\ref{phi_approx}), we obtain, for $\lambda > \lambda_c$,
\e
\phi_0({\bf x}, \lambda) = - c \tanh \bigg( \frac{\mu}{2} ({|{\bf x}|} - \lambda) \frac{\lambda}{\lambda_c} \bigg) =  - c \tanh \bigg( \frac{\mu}{2} \frac{({|{\bf x}|} - \lambda)}{\sqrt{1- {\dot\lambda}^2}}\bigg)  \,\,,
\q
so the classical path $\phi_0 ({\bf x}, \lambda)$ now describes an expanding nucleation bubble, as its physical radius $\lambda$ increases towards the limiting speed $\dot \lambda=1$,
with the proper Lorentz factor automatically included, and the effective tension now given by $S_1/{\sqrt{1- {\dot\lambda}^2}}$.

So we see that $\phi_0 ({\bf x}, \lambda)$ for real $\lambda$ spanning $\infty > \lambda > -\infty$ works equally well for classically allowed as well as classically forbidden regions.
In summary, $\phi_0 ({\bf x}, \lambda)=-c$ for $\lambda <0$, when $\phi$ stays in the false vacuum. For $\lambda_c \ge \lambda > 0$, $\phi_0 ({\bf x}, \lambda)$ describes the ``averaged" quantum fluctuation
that corresponds to the MPEP for the tunneling process under the potential barrier. The  $|{\bf x}| < \lambda$ region has fluctuated to the true vacuum, which is separated from the
false vacuum region by a domain wall at  $|{\bf x}| = \lambda$. Finally, for $\lambda \ge \lambda_c$, $\phi_0 ({\bf x}, \lambda)$ describes the classical propagation of the nucleation bubble.
The advantage here is that a single real parameter describes the whole system. The problem has been reduced to that of a time-independent one-dimensional (i.e., the $\lambda$
coordinate here) quantum mechanical system of a particle with position-dependent mass $m(\lambda)$ (\ref{m}) and a potential $U(\lambda)$ (\ref{Upot}) which has a barrier ($\lambda_c \ge \lambda > 0$)
that separates the two classically allowed regions. In general, the position-dependent mass complicates the quantization of the position variable. However, at leading order in the
WKB approximation, such a complication does not arise.

\section{Resonant Tunneling in Scalar Quantum Field Theory} \label{RTQFT}

The above discussion introduces the functional Schr\"odinger method and its applications to the tunneling process at leading order in $\hbar$. 
It reduces the tunneling process in an infinite-dimensional field configuration space to a one-dimensional
quantum mechanical tunneling problem.  The essential difference between tunneling in field theory discussed in Section \ref{functional_schroedinger} and in quantum mechanics
discussed in Section \ref{QM} is that in field theory one should first find the MPEP, namely $\phi_0({\bf x}, \lambda)$, and then obtain the effective potential $U(\lambda)$
(\ref{Upot}) and the effective mass $m(\lambda)$ (\ref{m}). We have extended the MPEP to include regions where classical motion is allowed.

\subsection{Setup}

\begin{figure}
\begin{center}
\includegraphics[width=0.8\textwidth]{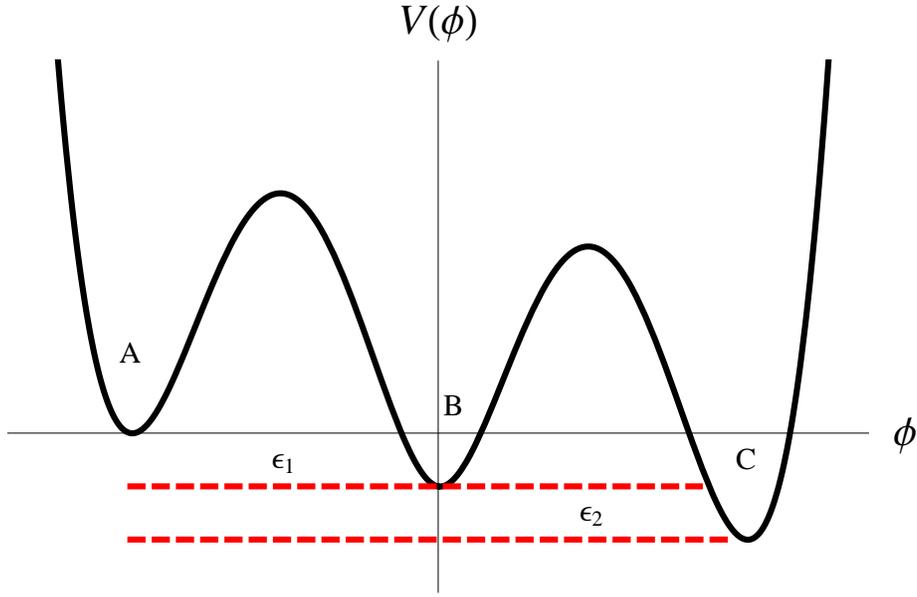}
\caption{The potential $V(\phi)$ in (\protect\ref{resonant_potential}). The vacuum $A$ is at $\phi_A = -2c_1$, with $V(\phi_A)=0$, the vacuum $B$ is at  $\phi_B = 0$, with $V(\phi_B)=-\epsilon_1$, and the vacuum $C$ is at $\phi_C = 2c_2$, with $V(\phi_C)= - \epsilon_1-\epsilon_2$.}
\label{V}
\end{center}
\end{figure}

To examine resonant tunneling in QFT, let us consider the following potential shown in Figure \ref{V},
\e
\label{resonant_potential}
V(\phi) = \left\{ \begin{array}{ll}
\frac{1}{4} g_1 ((\phi + c_1) ^ 2 - c_1 ^ 2) ^ 2 - B_1 \phi - 2 B_1 c_1 \quad \quad \phi < 0 \\
\frac{1}{4} g_2 ((\phi - c_2) ^ 2 - c_2 ^ 2) ^ 2 - B_2 \phi - 2 B_1 c_1 \quad \quad \phi > 0 
\end{array} \right.
\q
where as before $B_1$ and $B_2$ are small.  For this potential the false vacuum (vacuum A) at $\phi \approx -2c_1$ has zero energy density, the intermediate vacuum
(vacuum B) at $\phi = 0$ has an energy density $-\epsilon_1 = -2 B_1 c_1$ and the true vacuum (vacuum C) at $\phi \approx 2c_2$ has an energy density $-\epsilon_1-\epsilon_2 = -2
B_1 c_1- 2 B_2 c_2$.  We take both $\epsilon_1$ and $\epsilon_2$ to be small so that the thin-wall approximation is valid. Similar to the single barrier case, we introduce the inverse thickness $\mu_j= \sqrt{2 g_j c_j ^2}$ and the tension $S_1^{(j)}= 
\frac{2}{3}\mu_j c_j$ for each of the two domain walls: $j=1$ for the outside bubble and $j=2$ for the inside bubble. Here $r_1 > r_2$.
In the thin-wall approximation, $\phi=-2c_1$ for $r \gg r_1$, $\phi=0$ for $r_1 \gg r \gg r_2$, and  $\phi= +2c_2$ for $r \ll r_2$. 
The six parameters of the potential $g_{1,2}$, $c_{1,2}$ and $B_{1,2}$ now become $\mu_{1,2}$, $S_1^{(1,2)}$ and $\epsilon_{1,2}$.
In the thin-wall approximation, the thicknesses $1/\mu_{1,2}$ of the domain walls drop out, simplifying the discussion.

We also assume that the $O(4)$-symmetric solution provides the dominant contribution to the vacuum decay rate.  We note that the inside (half-)bubble does not have to be concentric with the outside (half-)bubble as long as the centers of the
two bubbles lie on the same spatial slice.  As we shall see, the analysis will go through without change as long as the two bubble walls are separated far enough, that is, much more than the
combined thicknesses $1/\mu_1 + 1/\mu_2$.  We expect the off-center bubble configurations to be subdominant if we include corrections to the thin-wall approximation.  We focus
here on the zero-energy (i.e., $E=0$) case.

\subsection{Ansatz}

The MPEP involves $\phi$ in the two under-the-barrier regions as well as the classically allowed region between them. In the under-the-barrier regions, we can solve for $\phi({\bf x},
\tau)$ using the Euclidean equation of motion (\ref{EL}), while in the classically allowed region, we can solve for $\phi({\bf x}, t)$ using the equation of motion in Minkowski space
(\ref{LEOM}). We can then convert them to $\phi_0({\bf x},\lambda)$.
 
However, in the thin-wall approximation, it is easier to simply write down the ansatz in the radial coordinate and then extract $\phi ({\bf x},\lambda)$ from it. Here the tunneling
process involves two concentric bubbles: an outside bubble whose domain wall separates $A$ (outside) from $B$ and an inside bubble whose domain wall separates $B$ from $C$ (inside).   
The radii of the two bubbles are $r_1$ and $r_2$ as shown in Figure \ref{bubbles}.
As a function of the four-dimensional radial coordinate $r$, we have the MPEP (see Appendix A),
\e
\label{gensol}
\phi(r) = -c_1 \tanh \bigg( \frac{\mu_1}{2} (r-r_1) \bigg) - c_2 \tanh \bigg( \frac{\mu_2}{2} (r-r_2) \bigg) +c_2-c_1
\q
For appropriate $r_1$ and $r_2$, this solves the Euclidean equation of motion (\ref{EL}) (and the Lorentzian equation of motion (\ref{LEOM}) in the appropriate regions). However,
here we shall use this $\phi$ (\ref{gensol}) as an ansatz to find the resonant tunneling condition. 
    
\begin{figure}
\begin{center}
\includegraphics[width=0.8\textwidth]{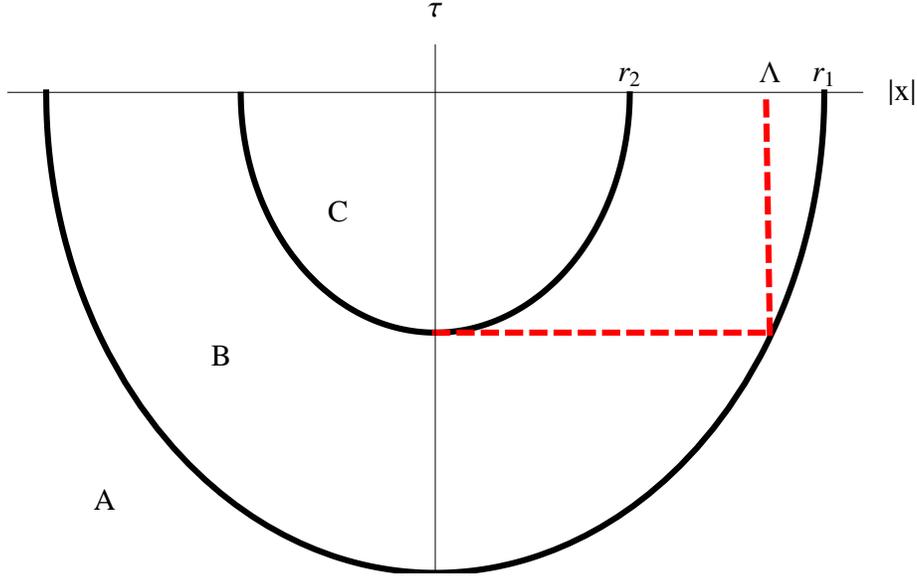}
\caption{The tunneling process from $A$ to $C$ via $B$ leads to the formation of two bubbles: the outside bubble separates $B$ from $A$ and the inside bubble separates $C$ from $B$. They are drawn as concentric bubbles here, though this is not the case in general. 
Here $\Lambda$ is the length of the horizontal dashed (red) line. We have $r_1 > \Lambda > r_2$.}
\label{bubbles}
\end{center}
\end{figure}

\begin{figure}
\begin{center}
\includegraphics[width=0.8\textwidth]{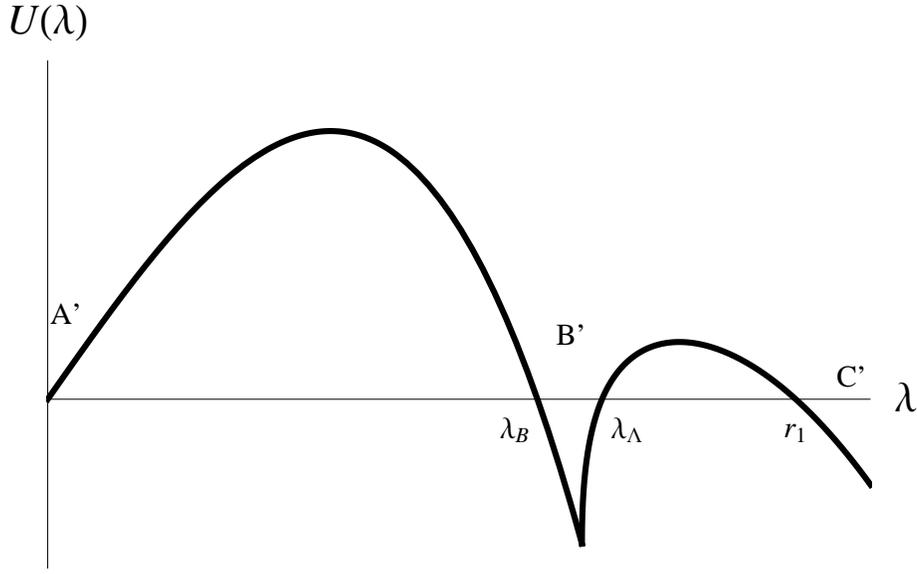}
\caption{A typical effective tunneling potential $U(\lambda)$ (\protect\ref{U_double}) for double tunneling.  The classical turning points for zero-energy tunneling are $0$, $\lambda_B$, $\lambda_{\Lambda}$, and $r_1$.  The
discontinuity in the derivative of $U(\lambda)$ occurs at $\lambda = \Lambda$. If the negative part of $U(\lambda)$ (i.e., the classically allowed region) is not too deep,
 $\lambda_B \gtrsim \lambda_{1c} = 3 S^{(1)}_1 / \epsilon_1$, and $\lambda_{\Lambda} \gtrsim \Lambda$.  In appropriate units, the values plotted here are $S_1^{(1)} = 1$,
 $S_1^{(2)} = 1.4$, $\epsilon_1 = 0.25$, $\epsilon_2 = 8.4 \cdot 10^{-3}$, $r_2 = 14$ which gives $r_1 = 20$ and $\Lambda = 14.3$ using (\protect\ref{geometry_constraint}) and (\protect\ref{action_constraint}).}
\label{double_barrier}
\end{center}
\end{figure}

Now it is straightforward to extract $\phi({|{\bf x}|}, \lambda)$ from $\phi (r)$ given by (\ref{gensol}),
\e
\label{MPEPRT}
\phi_0({|{\bf x}|},\lambda) = -c_1 \tanh \bigg( \frac{\mu_1}{2}\frac{\lambda}{r_1} ({|{\bf x}|}-\lambda) \bigg) 
-  c_2 \tanh \bigg( \frac{\mu_2}{2}\frac{\lambda'}{r_2} ({|{\bf x}|} - \lambda') \bigg) +c_2-c_1
\q
where we use the same reparametrization as in the single-barrier case. 
Here $\Lambda$ is the value of $\lambda$ at which the inside bubble has zero spatial extent,
\e
\label{geometry_constraint}
\Lambda ^2 = r_1 ^2 - r_2^2
\q
and as long as both bubbles are expanding
\e
\label{lambdaprime}
\lambda^{\prime} = \left\{ \begin{array}{ll}
\sqrt{\lambda ^2 - \Lambda^2} \quad \quad \quad \quad \Lambda < \lambda\\
0 \quad \quad \quad \quad \quad \quad \quad \quad \rm{otherwise}.
\end{array} \right.
\q
This is shown in Figure \ref{bubbles}. The equation (\ref{MPEPRT}) also implies that $\phi_0({|{\bf x}|}, \lambda)=-2c_1$ for $\lambda <0$.  Note also that the sum of the second
and third terms in (\ref{MPEPRT}) vanishes for $\lambda < \Lambda$.
Substituting this MPEP $\phi_0({|{\bf x}|}, \lambda)$ given by (\ref{MPEPRT}) into (\ref{Upot}) now yields, after a straightforward calculation, the effective tunneling potential
$U(\lambda)= U(\phi_0({|{\bf x}|},\lambda))$,
\e
\label{U_double}
U(\lambda) = 
2 \pi S_1^{(1)} \bigg( \frac{\lambda}{r_1} + \frac{r_1}{\lambda} \bigg) \lambda^2 - \frac{4 \pi}{3} \epsilon_1 \lambda^3
+ 2 \pi S_1^{(2)} \bigg( \frac{\lambda^ \prime }{r_2} + \frac{r_2}{ \lambda ^\prime} \bigg) (\lambda^\prime)^2 -
\frac{4 \pi}{3} \epsilon_2 (\lambda^ \prime)^3 
\q
For appropriate parameter choices, we see that (\ref{U_double}) has four zeros as shown in Figure \ref{double_barrier}. If the negative part of $U(\lambda)$ (i.e., the classically
allowed region) is not too deep, $\lambda_{\Lambda} \gtrsim 
\Lambda$, and $\lambda_B \gtrsim \lambda_{1c} = 3 S^{(1)}_1/ \epsilon_1$. We may take $U(\lambda)=0$ for $\lambda <0$. The
discontinuity in the derivative of $U(\lambda)$ at $\Lambda$ will be smoothed when the thickness of the bubble wall is taken into account.

\begin{figure}
\begin{center}
\includegraphics[width=0.8\textwidth]{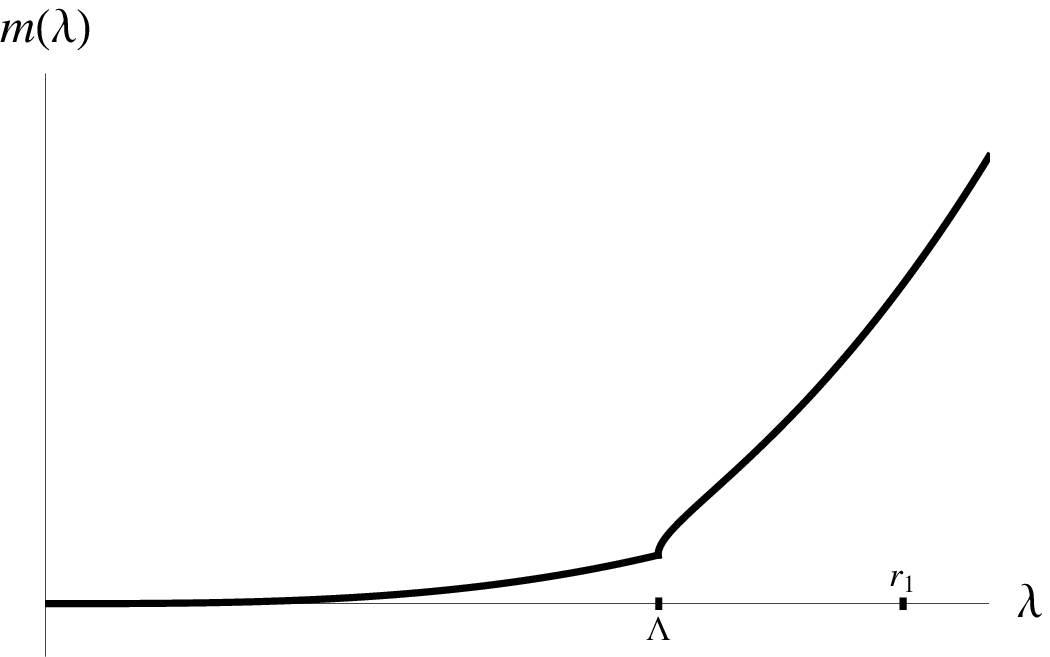}
\caption{The position-dependent mass $m(\lambda)$ (\protect\ref{m2}).  In appropriate units, the values plotted here are $S_1^{(1)} = 1$, $S_1^{(2)} = 1.4$, $\epsilon_1 = 0.25$, $\epsilon_2 = 8.4 \cdot 10^{-3}$.  The choice $r_2 = 14$ gives $r_1 = 20$ and $\Lambda = 14.3$ after solving the constraints (\protect\ref{geometry_constraint}) and (\protect\ref{action_constraint}).}
\label{m_double_barrier}
\end{center}
\end{figure}

It is also straightforward to evaluate the effective mass $m(\lambda)$ defined by (\ref{m}) using $\phi_0({|{\bf x}|},\lambda)$ of (\ref{MPEPRT}), now given by
\e
\label{m2}
m(\lambda) = 4 \pi \bigg( \frac{S_1^{(1)}}{r_1}\lambda^2 + \frac{S_1^{(2)} \lambda}{r_2 \lambda^
\prime}(\lambda^\prime)^2\bigg) \lambda \,\,.
\q
Note that, as expected, $m(\lambda) >0$.
Now we have a time-independent one-dimensional (with $\lambda$ as its coordinate) QM problem with the double-barrier potential $U(\lambda)$ (\ref{U_double}) and mass $m(\lambda)$
(\ref{m2}), which is illustrated in Figure \ref{m_double_barrier}.

\subsection{Constraints}

We see that the existence of the classically allowed region $B'$  in $U(\lambda)$ will lead to resonant tunneling.
We would like to see what properties of potential $V(\phi)$ (\ref{resonant_potential}) will yield resonant tunneling.
It is important to emphasize that the existence of the classically allowed region $B'$ is not guaranteed. 
For $E = 0$ the existence of a double-barrier $U(\lambda)$ potential requires four distinct classical turning points satisfying
\e
\label{energy_constraint}
0 = U(0)= U(\lambda_{B}) = U(\lambda_{\Lambda}) = U(r_1) \,\,.
\q
The radii of the two bubbles at the moment of nucleation are related via (\ref{geometry_constraint}).

We may evaluate the Euclidean action $S_E$ for $\phi$ (\ref{gensol}) in the thin-wall approximation. Minimization of $S_E$ with respect to $r_1$ and $r_2$ separately will
yield $r_1=\lambda_{1c}$ and  $r_2=\lambda_{2c}$ (see Appendix A). However, this bounce solution includes paths that passes through region $B$ only once. Since resonant tunneling must include
paths that bounce back and forth any number of times in the region $B$, this is not what we are seeking. Instead of finding the $S_E$ that includes multiple passes through $B$,
we use the the functional Schr\"odinger method to reduce the problem to a one-dimensional time-independent QM problem, which is then readily solved for $S_{(0)}$.

Once the simultaneous nucleation of the two bubbles is completed and just before they start to evolve classically, we are at $\lambda=r_1$, where $U(r_1)=0$, 
\e
\label{action_constraint}
\mathcal{E}_{(2)} = U(r_1) = 4 \pi (S_1^{(1)} - \frac{1}{3} r_1 \epsilon_1) r_1^2 + 4 \pi (S_1^{(2)} - \frac{1}{3}r_2 \epsilon_2) r_2^2 = 0
\q
This turns out to be the energy conservation condition as well. 
For the single-bubble case, the corresponding energy conservation condition (\ref{zeroE}) is equivalent to the minimization of the action. 
For the double-bubble case, the total energy $\mathcal{E}_{(2)}$ of the two bubbles at the moment of creation (at $\lambda= r_1$) 
must vanish. If we treat the region between the bubbles classically during the nucleation process, then the energy of the inside bubble,
that is, the second term in the above condition (\ref{action_constraint})  must vanish by itself (following from the condition (\ref{zeroE})), in which case the first term
vanishes as well. That is, $r_1=\lambda_{1c}=  3S_1^{(1)}/\epsilon_1$ and $r_2=\lambda_{2c}=  3S_1^{(2)}/\epsilon_2$.
However, it is crucial that the classically allowed region receives a full quantum treatment. So we must treat the simultaneous nucleation of the two bubbles quantum mechanically
and demand only $\mathcal{E}_{(2)}=0$. The requirement that the Euclidean action be stationary in this case reproduces (\ref{action_constraint}). See Appendix A for some details.

\begin{figure}
\begin{center}
\includegraphics[width=0.8\textwidth]{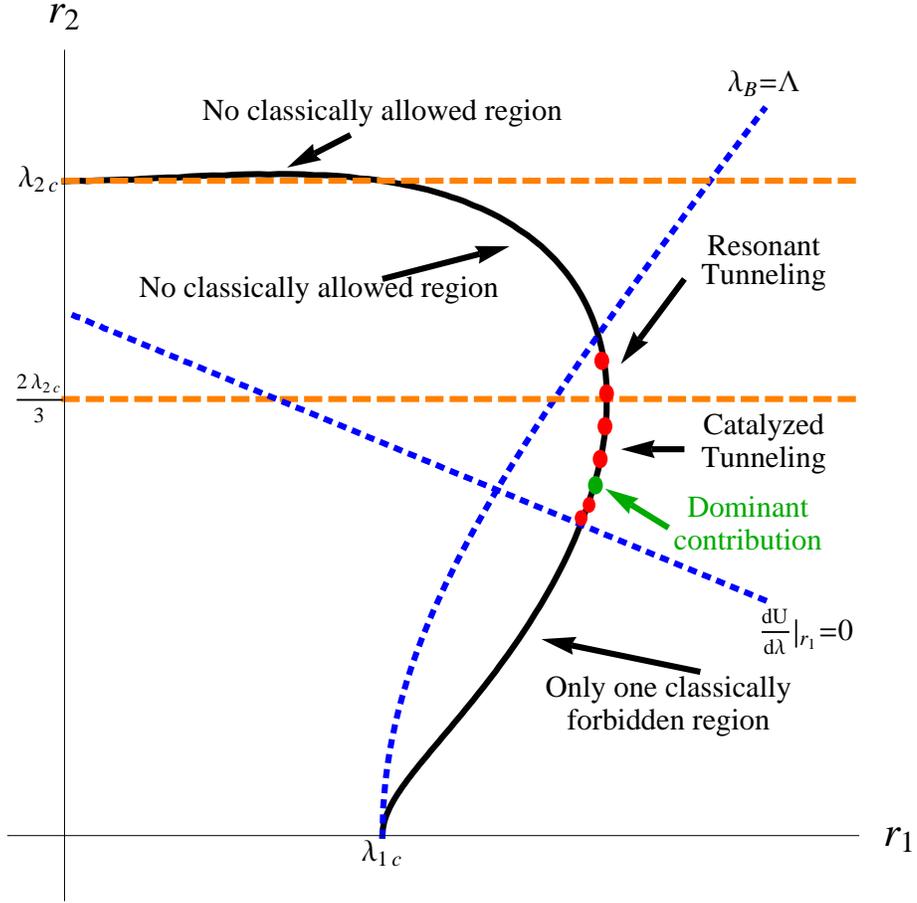}
\caption{The allowed parameter region for resonant tunneling. In appropriate units, the values plotted here are $S_1^{(1)} = 1$, $S_1^{(2)} = 5$, $\epsilon_1= 0.1$, $\epsilon_2 = 0.3$.  
The energy constraint (\protect\ref{action_constraint}) constrains $r_1$ and $r_2$ to lie on the black solid curve.  The region to the left of the blue dotted curve labeled $\lambda_B =
\Lambda$ is excluded since in this region $B^{\prime}$ does not exist.  The region below the blue dotted curve labeled $d U/ d \lambda|_{r_1} = 0$ is excluded since in this region
there is only one barrier in $U(\lambda)$. The red dots and green dot satisfy the Bohr-Sommerfeld quantization condition (\protect\ref{resonancecond}) in addition to satisfying the
consistency conditions (\protect\ref{action_constraint}), (\protect\ref{twobarriers}), (\protect\ref{r1r2constraint}), and (\protect\ref{completioncondition}).  The green dot with $r_2 \approx 0.54
\lambda_{2c}$ provides the dominant contribution to the tunneling probability.  Since this point lies below the orange dashed line, catalyzed tunneling occurs.  The tunneling
probability is $-\log T_{A \to B}^{\rm{res}} \approx 2.6 \cdot 10^4$, which is exponentially enhanced compared to the naive single-barrier tunneling probability $-\log T_{A
\to B}^{\rm{normal}} = \hat S_E \approx 1.3 \cdot 10^5$. }
\label{nconstraint}
\end{center}
\end{figure}

Note that the existence of a classically allowed region $B'$ implies that, $U(\lambda) < 0$ for $\Lambda > \lambda > \lambda_B$.
Following (\ref{U_double}), we obtain 
\e
\label{twobarriers}
\Lambda^2 >  \lambda_{B}^2 = \frac{\lambda_{1c} r_1^2}{2 r_1-\lambda_{1c}}
\q
from which it follows that  ($\Lambda^2=r_1^2 -r_2^2$)
\ba
\label{r1r2constraint}
r_1 &>& r_2 \\ \nonumber
r_1 &>& \lambda_{B}  > \lambda_{1c} = 3S_1^{(1)}/\epsilon_1\\ \nonumber
r_2 &<& \lambda_{2c} = 3S_1^{(2)}/\epsilon_2 \,\,.
\ea  
The existence of a second classically forbidden region requires $\lim_{\lambda \to r_1^-} d U/ d \lambda|_{\lambda} < 0$ since $U(r_1)=0$ is automatically
satisfied.  Equivalently
\e
\label{completioncondition}
2(S_1^{(1)} + S_1^{(2)}) < r_1 \epsilon_1 + r_2 \epsilon_2 \,\,.
\q
These constraints are illustrated in Figure \ref{nconstraint}. 
We see that the simultaneous nucleation of two bubbles can now be parameterised by a single parameter, say $r_2$. 
When permitted, the sizes of the bubbles, namely $r_1$ and $r_2$, will be such that the resonance condition $W = (n+1/2) \pi$ is satisfied.

\begin{figure*}
  \centerline{
    \mbox{\includegraphics[width=0.4\textwidth]{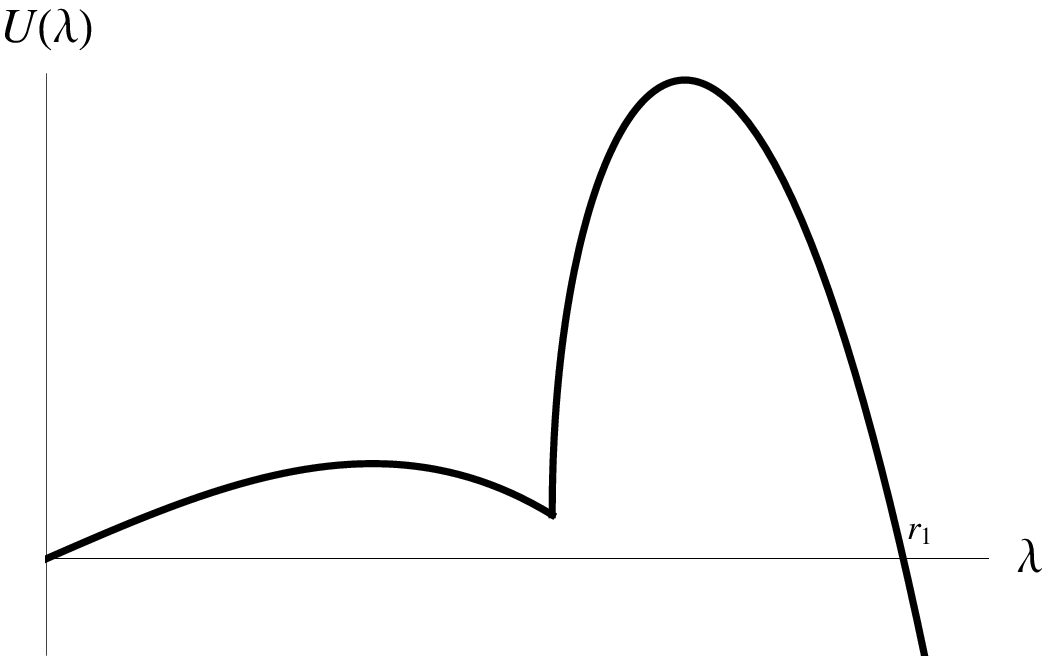}}
    \mbox{\includegraphics[width=0.4\textwidth]{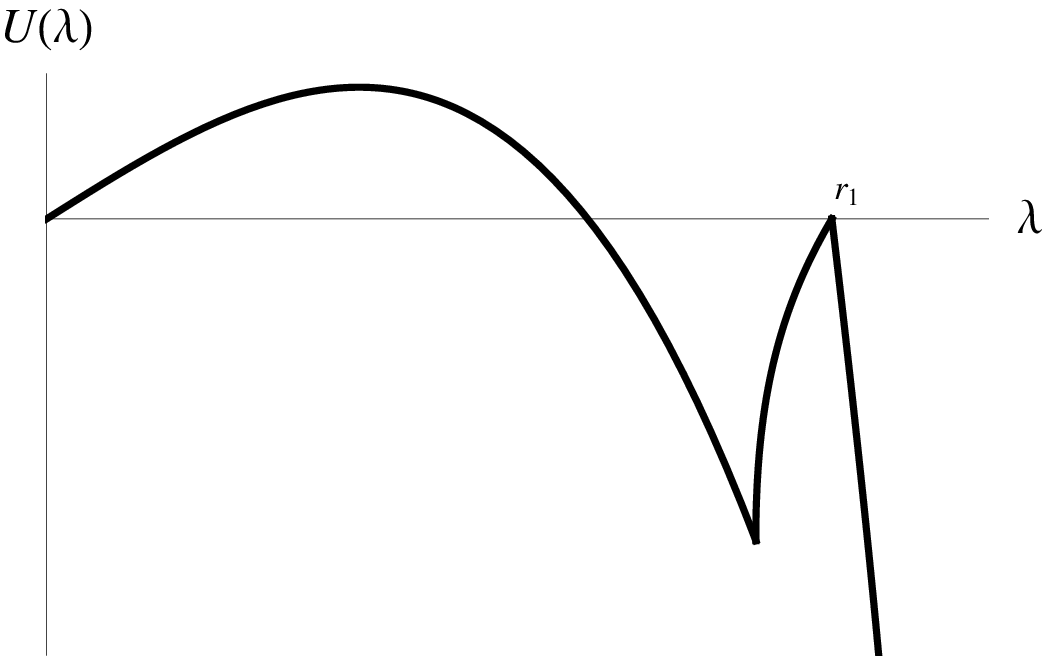}}
    }
  \caption{Two examples where resonant tunneling is absent. In appropriate units, the values plotted here are $S_1^{(1)} = 1$, $S_1^{(2)} = 5$, $\epsilon_1= 0.1$, $\epsilon_2 = 0.3$.  The left plot shows the effective tunneling
  potential $U(\lambda)$ for $r_2 = 0.8 \lambda_{2c} > r_{2,\rm{max}}$.  No classically allowed region exists for this potential.  The right plot shows the effective tunneling
  potential for $r_2 = 0.4 \lambda_{2c} < r_{2,\rm{min}}$.  There is no second barrier for this potential.  The discontinuities in the derivative of $U(\lambda)$ 
  will be smoothed out when the thickness of the bubble walls are taken into account.}
\label{excluded}
\end{figure*}

The width of the classically allowed region $\Delta \lambda_{B^\prime} \equiv \lambda_{\Lambda} - \lambda_B$ in $U(\lambda)$ decreases monotonically as $r_2$ increases.  The
classically allowed region is a point when $\Delta \lambda_{B^\prime} =0$ at some maximum value $r_{2,\rm{max}}$ of $r_2$ (when (\ref{twobarriers}) is saturated).  When $r_2 >
r_{2,\rm{max}}$, there is no classically allowed region in $U(\lambda)$.  Similarly the width of the second classically forbidden region $\Delta \lambda_{\rm{barrier}} \equiv r_1
- \lambda_{\Lambda}$ increases monotonically as $r_2$ increases.  At some minimum value $r_{2,\rm{min}}$ of $r_2$ (\ref{completioncondition}) is saturated, and the second barrier
in $U(\lambda)$ becomes a single point.  The condition that $\Delta \lambda_{\rm{barrier}} > 0$ is equivalent to the condition $\lim_{\lambda \to r_1^-} d U/ d \lambda|_{\lambda}
< 0$.  Figure \ref{excluded} shows a typical effective tunneling potential in each of these two cases.

After the simultaneous nucleation of the two bubbles quantum mechanically, the outside bubble will grow so the tunneling out of vacuum $A$ will complete. Now there are two
possibilities for the inside bubble, depending on whether it has the critical size (\ref{Rgrow}) to grow or not (note that the binding energy of the two bubbles is expected to be negligible):

(1) $\lambda_{2c}  >  r_2 > 2 \lambda_{2c}/3$, in which case the inside bubble will grow as well. Hence the tunneling from vacuum $A$ to vacuum $C$ will complete, although the outside bubble is expected to grow faster than the inside bubble. This is the analogue of resonant tunneling in quantum mechanics, so we refer to this tunneling process from $A$ to $C$ via $B$ as resonant tunneling when $W = (n+1/2) \pi$.

(2) $0 <  r_2 < 2 \lambda_{2c}/3$, in which case the inside bubble will collapse after nucleation, while the outside bubble will grow. In this case, the tunneling from $A$ to $B$
will complete. At a later time, tunneling from $B$ to $C$ will take place via a normal tunneling process. In this process, the presence of vacuum $C$ can increase the tunneling
rate from $A$ to $B$ by an exponential factor compared to the naive rate given by (\ref{S_E0}). We refer to this tunneling process from $A$ to $B$ in the presence of vacuum $C$ as
assisted or catalyzed tunneling, since $C$ plays the role of a catalyst.  Note that in this region (\ref{lambdaprime}) is modified for $\lambda > r_1$:
\e
\label{lambdaprimenew}
\lambda^{\prime} = \left\{ \begin{array}{ll}
\sqrt{\lambda ^2 - \Lambda^2} \quad \quad \quad \quad \Lambda < \lambda \leq r_1\\
\sqrt{r_1^2+r_2^2 - \lambda^2} \quad \quad r_1 < \lambda < \sqrt{r_1^2 +r_2^2} \\
0 \quad \quad \quad \quad \quad \quad \quad \quad \rm{otherwise}.
\end{array} \right. 
\q
The inside bubble shrinks for $\lambda > r_1$ and disappears entirely when $\lambda = \sqrt{r_1^2 +r_2^2}$.

\subsection{Tunneling Probability}

Explicitly, as shown in Figure \ref{double_barrier}, the classical turning points (with $E=0$) are $0$, $\lambda_B$, $\lambda_{\Lambda}$ and $r_1$. 
Then the single-barrier tunneling probability is given by
\e
T_{A' \to B'} \simeq \frac{4}{\Theta ^2} \,\,,
\q
where now
\e
\label{theta}
\Theta = 2 \exp \bigg( \int _{0}^{\lambda_B} d \lambda \sqrt{2 m(\lambda) U(\lambda)}\bigg) \,\,.
\q
In the thin-wall limit with $E=0$
\e
\label{log_theta}
\ln \bigg( \frac{\Theta}{2}\bigg) = \frac{\pi^2}{4} \lambda_B^3 S_1^{(1)} 
\q
As before, the tunneling probability (\ref{T_AtoB}) calculated using the
functional Schr\"odinger equation, with $\Theta$ given by (\ref{log_theta}), agrees with the result of the Euclidean instanton method, $\exp(-S_E / \hbar)$, where $S_E$ is given
by (\ref{S_E0}).  

The tunneling probability from vacuum $A'$ to vacuum $C'$ via the intermediate vacuum $B'$, is given by
(\ref{A_to_C}) where now $W$ and $\Phi$ are given by
\ba
\label{W}
W &=& \int_{\lambda_B}^{\lambda_{\Lambda}} d \lambda \sqrt{2 m(\lambda) ( - U(\lambda))} \nonumber \\
 &=&  \frac{S_1^{(1)} \lambda_{\Lambda}}{\lambda_B} \sqrt{\lambda_{\Lambda}^2 - \lambda_B^2} - S_1^{(1)} \lambda_B \log \bigg[ \frac{\lambda_{\Lambda} + \sqrt{\lambda_{\Lambda}^2 -\lambda_B^2}}{\lambda_B}\bigg] \nonumber \\
&\approx & \frac{S_1^{(1)} \Lambda}{\lambda_{1c}} \sqrt{\Lambda^2 - \lambda_{1c}^2} - S_1^{(1)} \lambda_{1c} \log \bigg[ \frac{\Lambda + \sqrt{\Lambda^2 -\lambda_{1c}^2}}{\lambda_{1c}}\bigg]  \,\,.
 \ea
with the classical turning points shown in Figure \ref{double_barrier}.  The third equality in (\ref{W}) is valid if the classically allowed region is shallow. In this approximation, we also have, with $\lambda_{\Lambda} \approx \Lambda$,
\ba
\ln \bigg( \frac{\Phi}{2}\bigg) &=& \int_{\lambda_{\Lambda}}^{r_1} d \lambda \sqrt{2 m(\lambda) U(\lambda)} \\ \nonumber
&\approx & \frac{S_E}{2} -  \frac{\pi^2}{4} \Lambda^3 S_1^{(1)}
\ea
where $S_E$ (\ref{SE1}) is given in the appendix. Since $\Lambda$, $r_1$ and $r_2$ are related by (\ref{geometry_constraint}) and (\ref{action_constraint}), we may consider $\Theta(r_2)$, $W(r_2)$ and $\Phi(r_2)$ as functions of $r_2$ only.

The bubble sizes are dominated by the ones that satisfy the resonance condition (\ref{resonancecond}), i.e., 
$W=(n_{B}+1/2) \pi$ for the $n_B$th resonance. With $r_2$ satisfying this condition and the constraints shown in Figure \ref{nconstraint}, 
the resulting tunneling probability is now given by (\ref{resonancetnn}):
$$T_{A \rightarrow C}  = \frac{4}{\left(\Theta/\Phi+ \Phi/\Theta \right)^{2}}$$
which can approach unity for suitably chosen potential (\ref{resonant_potential}).

Next let us consider catalyzed tunneling. This is the case when the inside bubble classically re-collapses after its creation. The normal probability $T_{A \rightarrow B}$ has a bounce value ${\hat S_E}$ smaller than that given by (\ref{log_theta}), or
\e
T^{\rm{normal}}_{A \rightarrow B} \simeq e^{-{\hat S_E}} = e^{- \pi^2\lambda_{1c}^3 S_1^{(1)}/2} >  4/ {\Theta}^2 
\q
where $\Theta$ is given by (\ref{log_theta}).
The presence of vacuum $C$ can lead to an enhanced tunneling probability,
$$T^{\rm{res}}_{A \rightarrow B}  \approx \frac{4}{\left(\Theta/\Phi+ \Phi/\Theta \right)^{2}}$$
which can be substantially bigger than $T^{\rm{normal}}_{A \rightarrow B}$ if $\Theta \sim \Phi$. On the other hand, the catalytic effect is negligible if $\Phi$ is exponentially too big or too small when compared to $\Theta$. 

\subsection{Generic Situation} \label{Discussion}

For large $\Theta$ and $\Phi$, so that the penetration through the barriers
is strongly suppressed, the tunneling probability has sharp
narrow resonance peaks at the values in (\ref{resonancecond}). 
If we allow the possibility of a non-zero inital energy $E$ that is small compared to all other relevant mass scales, we introduce an extra variable without introducing any additional constraints, although (\ref{energy_constraint}) and (\ref{action_constraint}) will be slightly modified.  
Treating the resonance shape as a function of energy $E$,
the resonance has a width $\Gamma_{E}$. Expanding around the resonance at $E=E_R$, we have
\be
\cos W = \pm \left(\frac{\partial W}{\partial E} \right) {\bigg|}_{E_R} (E-E_R), \quad \quad \sin W =1
\ee
and  
\be
T_{A \rightarrow C}  \propto  \frac{1}{(E-E_R)^2 + (\Gamma_{E}/2)^2}
\ee 
so this yields, for large $\Theta$ and $\Phi$,
\be
\Gamma_{E}= \frac{2}{\Theta \Phi (\frac{\partial W}{\partial E})} \left(\frac{\Theta}{\Phi} + \frac{\Phi}{\Theta} \right)
\ee
Next, let the separation between neighboring resonances be $\Delta E$, where
\be
\Delta E \simeq \frac{\pi}{(\frac{\partial W}{\partial E})} 
\ee
Then a good estimate of the probability of hitting a resonance is given by
\ba
P(A \rightarrow C) =\frac{\Gamma_{E}}{\Delta E} \simeq \frac{2}{\pi \Theta \Phi} 
\left(\frac{\Theta}{\Phi} + \frac{\Phi}{\Theta} \right)
= \frac{1}{2\pi} \left(T_{A \rightarrow B}  + T_{B \rightarrow C} \right)
\label{2steps}
\ea
We see that the probability of hitting a resonance is given by the larger of the two decay probabilities,
$T_{A \rightarrow B}$ or $T_{B \rightarrow C}$, and the average tunneling probability is given by 
\ba
<T_{A \rightarrow C}> = P(A \rightarrow C)T_{A \rightarrow C} 
\sim \frac{T_{A \rightarrow B}T_{B \rightarrow C}}{T_{A \rightarrow B} + T_{B \rightarrow C}}
\label{averageT2}
\ea
which is essentially given by the smaller of the two tunneling probabilities. This is a derivation of Eq.(\ref{gammaAtoC}).
Following the argument for (\ref{tAtoC}), the generalization to tunneling with multiple barriers is straightforward.

\section{Remarks} \label{Conclusion}

Our results do not conflict with the no-go theorem of \cite{Copeland} since the assumptions are inapplicable as anticipated by \cite{Sarangi:2007jb}.  (Note that in
\cite{Copeland} the term MPEP is used exclusively to refer to the path in the classically forbidden region; our more inclusive definition will be used in the discussion here.)  In
particular one of the assumptions is that the MPEP is everywhere stationary $\frac{\partial\phi_0}{\partial \tau} = 0$ or $\frac{\partial\phi_0}{\partial t} = 0$ at the boundary
between the classically allowed region and the classically forbidden region, i.e. for $\lambda = \lambda_B$ and $\lambda = \lambda_{\Lambda}$.  This condition is clearly violated by our
MPEP (\ref{MPEPRT}) which is only stationary everywhere for $\lambda \leq 0$ and $\lambda = r_1$.  The exact solution (\ref{gensol}) also violates this condition.  

In \cite{Saffin:2008vi} it was shown how a bubble of vacuum A surrounded by vacuum B could produce a bubble of vacuum C with probability of order unity under certain conditions. 
It was assumed that the vacuum energy density of vacuum C was greater than the vacuum energy density of vacuum A or vacuum B.  The inhomogeneous initial state violates one of the
conditions of the no-go theorem.  The physics is quite different because the asymptotic false vacuum is intermediate in field space. Since our analysis applies only in the double
thin-wall approximation, it is possible that the oscillons or its quantized version may play an important role in a different setup.

We apply the functional Schr\"odinger method to show how resonant tunneling takes place in quantum field theory with a single scalar field. Our analysis is carried out in the
double thin-wall approximation. The double-barrier potential problem in QFT is reduced to a double-barrier potential problem in a time-independent one-dimensional QM problem, so
the quantum mechanical analysis can be applied. 

The relevance of resonant tunneling is obvious if the potential has many local minima, as is the case of the cosmic landscape in string theory. So resonant tunneling in the
presence of gravity is a very important question to be addressed. 

What happens if the conditions (\ref{action_constraint}), (\ref{twobarriers}), (\ref{r1r2constraint}), and (\ref{completioncondition}) cannot be satisfied? Generically, the double
thin-wall approximation breaks down and a more careful analysis is needed. Based on the analysis of (\ref{tAtoC}), we are led to believe that the resonant tunneling phenomenon
will continue to happen. So it is interesting to study more general cases to obtain a complete picture. 

\appendix
\section{Euclidean Bounce Solution}

Recall that, for a O(4)-invariant bounce, with $r^2= \tau^2 + |{\bf x}|^2$, the Euclidean equation of motion becomes
\e
\label{Euceomc}
\frac{d^2 \phi}{d r^2} + \frac{3}{r} \frac{d \phi}{d r}  =  \frac{d V_E(\phi)}{d \phi}
\q
where the Euclidean potential $V_E (\phi) =-V(\phi)$ and 
\e
\label{conda1}
\lim_{r \to\infty} \phi(r)=\phi_A \equiv -2c_1
\q
Also,
\e
\label{conda2}
\frac{\partial \phi}{\partial r}\bigg|_{r=0} =0
\q
To demonstrate that a solution always exists, we simply follow Coleman's argument \cite{Coleman:1977py} that there are undershoot solutions as well as overshoot solutions.
Continuity then implies the existence of the solution. If we start with $\phi(0) < \phi_s$ where $\phi_s$ is close to but on the left of $\phi_C \equiv 2c_2$, where
$V_E(\phi_s)=V_E(\phi_B)=V_E(0)$ (that is $V_E(\phi(0))< V_E(\phi_B)$), then because of the damping term, we shall undershoot.
For overshoot, we start at $\phi_t$ very close to $\phi_C$, so we can linearize (\ref{Euceomc}),  
\e
\left( \frac{d^2}{d r^2} + \frac{3}{r} \frac{d }{d r} - \mu^2 \right) (\phi-\phi_s) = 0
\q
where $\mu^2= V_E''(\phi_C)$.
The solution can be expressed in terms of a Bessel function,
\e
\phi(r)-\phi_s = 2[\phi_s-\phi_C] I_1(\mu r)/\mu r
\q
where $\phi_s-\phi_C<0$ so $\phi$ is moving to the left. For $\phi_s$ sufficiently close to $\phi_C$, we can arrange for $\phi$ to stay arbitrarily close to $\phi_C$ for arbitrarily large $r$. For large enough $r$, the damping term is negligible. Without damping, $\phi$ can simply move beyond $\phi_A$. 

To satisfy the boundary conditions (\ref{conda1},\ref{conda2}), we expect a unique solution. We may evaluate the Euclidean action $S_E$ and then minimize it, which is equivalent to solving the above system (\ref{Euceomc}-\ref{conda2}). 
In the thin-wall approximation, the solution will take the form given by (\ref{gensol}). This yields
\e\label{SE1}
S_E = 2 \pi^2r_1^3 S^{(1)}_1 - \frac{\pi^2}{2} \epsilon_1(r_1^4-r_2^4) +2 \pi^2r_2^3 S^{(2)}_1 - \frac{\pi^2}{2} (\epsilon_1+\epsilon_2) r_2^4 
\q
A simple minimization of the resulting $S_E$ with respect to $r_1$ and $r_2$ separately will yield 
\e
\frac{\delta S_E}{\delta r_j} =0 \to r_j=\lambda_{jc}
\q
However, this bounce solution includes paths that passes through region $B$ only once.
It is crucial to recognize that possible resonant tunneling must include, in terms of Feynmann's path integral formalism, paths that bounce back and forth any number of times in the region $B$. So, we are not interested in
minimizing the action with a single pass through $B$. Instead, we should leave  $r_1$ and $r_2$ free at this moment. Once the simultaneous nucleation of the two bubbles is
completed and just before they start to evolve classically, energy conservation demands the condition (\ref{action_constraint}). For the single-bubble case, the corresponding
energy conservation condition (\ref{zeroE}) is equivalent to the minimization of the action. For the double-bubble case, we can obtain the condition (\ref{action_constraint}) by varying $S_E$ with respect to $dr_1=dr_2$.   

Instead of finding $S_E$ that includes multiple passes through $B$, we use the functional Schr\"odinger method to reduce the problem to a one-dimensional time-independent QM problem, which is then readily solved. Here,  the energy conservation condition (\ref{action_constraint}) is simply the turning point of $U(\lambda)$.

\vspace{0.6cm}

\noindent {\bf Acknowledgments}

\vspace{0.3cm}

We thank Shau-Jin Chang, Xingang Chen, Ed Copeland, Kurt Gottfried, Tony Padilla, Paul Saffin, Sash Sarangi, Gary Shiu, Ben Shlaer, Gang Xu and Yang Zhang for valuable discussions.
This work is supported by the National Science Foundation under grant PHY-0355005.

\vspace{0.5cm}

\end{document}